%% file: qso_reion_rev.tex
\documentclass[useAMS,usenatbib,usegraphicx]{mn2e}
\usepackage{graphicx}
\usepackage[latin1]{inputenc}
\usepackage{color}
\usepackage{times}
\usepackage{natbib}
\usepackage{setspace}
\newif\ifAMStwofonts
\AMStwofontstrue
\definecolor{red}{rgb}{1,0.,0.}

\def\lesssim{\lower.5ex\hbox{$\; \buildrel < \over \sim \;$}}
\def\gtrsim{\lower.5ex\hbox{$\; \buildrel > \over \sim \;$}}
\title[Reionization Sources] {On the relative Contribution of high-redshift Galaxies and Active Galactic Nuclei to Reionization.}

\author[Fontanot et al.]{
  \parbox[t]{\textwidth}{Fabio Fontanot$^{1,2,3}$, Stefano Cristiani$^3$
    \& Eros Vanzella$^3$}
    \vspace*{6pt}\\
    $^1$ HITS-Heidelberger Institut f\"ur Theoretische Studien, Schloss-Wolfsbru
nnenweg 35, 69118 Heidelberg, Germany\\
    $^2$ Institut f\"ur Theoretische Physik, Philosophenweg, 16, 69120, Heidelberg, Germany \\
    $^3$ INAF-Osservatorio Astronomico di Trieste, Via Tiepolo 11, I-34131 Trieste, Italy \\
    email: fabio.fontanot@h-its.org}

\begin{document}
\date{Accepted ... Received ...}

\maketitle

\begin{abstract}
In this paper we discuss the contribution of different astrophysical
sources to the ionization of neutral hydrogen at different redshifts.
We critically revise the arguments in favour/against a substantial
contribution of Active Galactic Nuclei (AGNs) and/or Lyman Break
Galaxies (LBGs) to the reionization of the Universe at $z>5$.  We
consider extrapolations of the high-z QSO and LBG luminosity functions
and their redshift evolution as well as indirect constraints on the
space density of lower luminosity Active Galactic Nuclei based on the
galaxy stellar mass function. Since the hypothesis of a reionization
due to LBGs alone requires a significant contribution of faint dwarf
galaxies and a LyC photon escape fraction ($f_{\rm esc}$) of the order
of $\sim 20\%$, in tension with present observational constraints, we
examine under which hypothesis AGNs and LBGs may provide a combined
relevant contribution to the reionization. We show that a relatively
steep faint-end of the AGN luminosity function, consistent with
present constraints, provides a relevant (although sub-dominant)
contribution, thus allowing us to recover the required ionizing photon
rates with $f_{\rm esc}\sim 5\%$ up to $z\sim7$. At higher redshifts,
we test the case for a luminosity-dependent $f_{\rm esc}$ scenario and
we conclude that, if the observed LBGs are indeed characterized by
very low $f_{\rm esc}$, values of the order of $f_{\rm esc} \sim 70\%$
are needed for objects below our detection threshold, for this galaxy
population to provide a substantial contribution to
reionization. Clearly, the study of the properties of faint sources
(both AGNs and LBGs) is crucial.
\end{abstract}


\section{Introduction}\label{sec:intro}
Cosmic reionization is a major focus in present Cosmology, since it
represents a crucial cosmic epoch for the formation of the first
structures and the production of the photons responsible for the end
the Dark post-recombination Ages. Moreover, these energetic photons
affect, in a critical interplay, the species available for gas cooling
(and consequently the star formation) and the collapse of (small) dark
matter halos.

An important constraint on the epoch when this phase transition occurs
is set by the measurement of the Thomson optical depth of the
intergalactic medium (IGM) via the large scale polarization of the
Cosmic Microwave Background (CMB, \citealt{WMAP7} ), which provides -
with the rough assumption of instantaneous reionization - $z_{reion} =
10.6 \pm 1.2$. Additional evidence comes from the Gunn-Peterson test
applied at the Lyman-$\alpha$ forest, characterized by a low neutral
hydrogen fraction at redshifts below 6 (\citealt{Fan06}, but see also
\citealt{McGreer11}). Recent evidence for a damping wing around the
systemic redshift of a z=7.085 quasar \citep{Bolton11} and for a
sudden decrease in the fraction of Lyman-$\alpha$ emitters among
$z\sim 7$ Lyman-break galaxies (LBG, see e.g., \citealt{Pentericci11,
  Schenker12, Ono12}) also suggests a rapid increase of the neutral
fraction of hydrogen in the Universe at these epochs. All these
evidences broadly constrain the epoch of hydrogen reionization in the
redshift range $6<z<12$.

At the same time, many studies have addressed the nature of the
ionizing sources. Pop III stars (e.g. \citealt{Ciardi00}),
star-forming galaxies (e.g. \citealt{Robertson10}), and Active
Galactic Nuclei (AGNs, e.g. \citealt{Haiman98}) have found to be prime
candidates, with more exotic possibilities explored in the form of
primordial black-holes and mini-quasars (e.g. \citealt{Madau04}), and
decaying particles (dark matter and neutrinos, e.g. \citealt{Scott91,
  Pierpaoli04}). It is commonplace that by $z \sim 6$ the ionizing
radiation emitted by quasars alone is insufficient to reionize the IGM
(e.g. \citealt{SchirberBullock03, Cowie09}); on the other hand
high-redshift galaxies are in principle able to produce the bulk of
the cosmic emissivity ionizing the IGM (see \citet{HaardtMadau12} for
a recent analysis), but only if the fraction of the 1-4 Ryd photons
escaping the galaxies is significant ($ \sim 20\%$ at $z\simeq 7$, see
also \citealt{Bouwens11b}) and if the contribution of very faint,
undetected objects is taken into account, assuming their space density
is satisfactory described by the extrapolation of the faint-end of
observed luminosity function (LF). Nonetheless, the direct detection
of Lyman continuum (LyC) photons from low redshift ($z<1.5$) galaxies
has been so far unsuccessful \citep{Cowie09,Siana10}, down to $f_{esc}
< 1\%$, and the measurements at the highest redshifts for which the
direct LyC measurement is still allowed by the rapidly increasing IGM
opacity ($z = 3-4$) have been shown to be prone to a significant
degree of contamination by lower-redshift interlopers
\citep{Vanzella10a, Vanzella12}.

The results of numerical hydrodynamical simulations (see
e.g. \citealt{Ciardi12}) also favour sources with a soft spectral
energy distribution (indicative of Population II stars) and high
escape fractions as the main contributors to Hydrogen reionization. On
the other hand, numerical simulations have not provided yet a
concordant answer to the question of $f_{esc}$ evolution as a function
of galaxy properties, luminosities and redshift: some groups reported
evidence for a {\it decrease} of $f_{\rm esc}$ with decreasing halo
mass at $z>3$ (e.g. \citealt{Gnedin08}), while competing groups have
found the exactly opposite result of an {\it increase} of $f_{\rm
  esc}$ with decreasing halo mass (e.g. \citealt{Yajima11}, with a
large scatter in the individual $f_{\rm esc}$ determinations).

The magnitude of the faintest dwarf galaxies represents another
critical point in the analysis. The expected absolute magnitude at
$145$ nm ($M_{\rm UV}$) of a galaxy hosted in a halo with virial
temperature $\sim 2 \times 10^4 K$ is $M_{\rm UV} \sim -10$ (see e.g,
\citealt{Trenti10}). However, theoretical models also suggest that
star formation might be strongly suppressed in early $M_{\rm UV} \gtrsim
-13$ dwarfs due to metallicity effects \citep{KrumholzDekel11,
  Kuhlen11}. Moreover, if faint galaxies were the major contributors
to reionization, this would imply an extended reionization epoch, a
result disfavoured by recent constraints on kinetic Sunyaev-Zel'dovich
effect \citep{Zahn11, Kuhlen12}. Therefore, if these dwarfs would not
be able to provide the required contribution to the ionizing
background, additional assumptions have to be tested, as a strong
luminosity/redshift evolution of $f_{\rm esc}$ or the inclusion
additional sources of ionizing photons.

All these considerations point out the need of a deeper investigation
of the (possibly sub-dominant) contribution of AGNs to the ionizing
flux, in order to explore to which extent these sources may be able to
alleviate the tensions arising from considering galaxies as the only
contributors to cosmic reionization. In this paper, we thus revisit
and compare the relative contribution of AGNs and galaxies to the
ionizing background at $z>5$. In particular, we focus on the putative
role played by faint AGNs, up to the same magnitude limits considered
for galaxies. Both the determination of the faint end of the observed
AGN luminosity function and the reliability of its extrapolation to
fainter magnitudes present similar challenges with respect to
analogous statistical estimators for galaxies. An additional source of
uncertainty, however, arises from the apparently discrepant results
obtained using independent selection techniques for the reference
sample, i.e. X-ray or optically based, which show a different
sensitivity to the various AGN populations. Nonetheless, it is also
possible to combine the most recent determination the galaxy stellar
mass function of galaxies \citep{Santini12} with empirical arguments,
to get an estimate of the maximum contribution of the AGN population
to the cosmic reionization.

The structure of this paper is as follows: in sec.~\ref{sec:models} we
present the formalism we adopt to estimate the contribution of each
class of sources with respect to the required ionization photon rate;
while in sec.~\ref{sec:results} we discuss the implication of our
finding and in sec.~\ref{sec:concl} we present our
conclusions. Throughout this paper we assume that QSOs represent the
luminous sub-population of the homogeneous AGN population, and that
Lyman Break Galaxies are a good tracer of the overall galactic
population.

\section{Modeling the contribution to reionization of different astrophysical sources}
\label{sec:models}

\subsection{High-z QSO/AGN Luminosity Functions} \label{subsec:QSO}

\begin{figure}
\includegraphics[width=9cm]{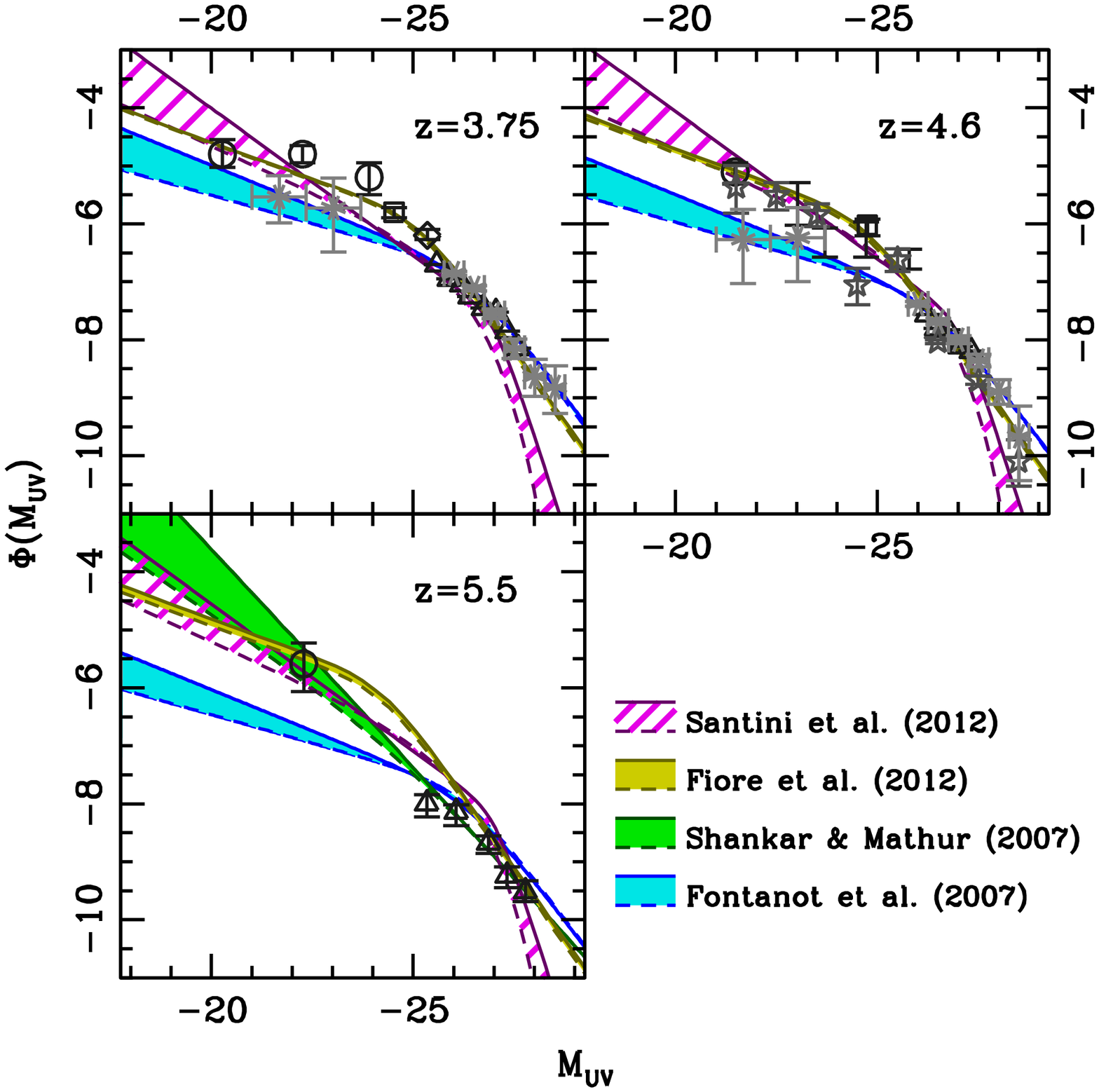}
  \caption{Comparison of different estimates for the high-z
    QSO-LF. Blue lines and light blue shaded region correspond to the
    best fit models in \citet{Fontanot07a}, and their redshift
    evolution. Green lines and light green shading correspond to the
    models in \citet{ShankarMathur07}. Yellow line corresponds to the
    evolution of X-ray selected QSO-LF in \citet{Fiore11}. Red lines
    and hatched light red shading correspond to the GSMF-derived
    high-z AGN-LF (see text for more details on the conversion from
    GSMF to QSO-LF). Dots refer to datapoints from \citet[grey
      asterisks]{Fontanot07a}, \citet[open circles]{Fiore11},
    \citet[open squares]{Civano11}, \citet[open diamonds]{Brusa10},
    \citet[stars]{Glikman11}, \citet[open triangles]{Jiang09}.  }
\label{fig:qsolf}
 \end{figure}

In order to estimate the contribution of quasars (QSOs) to the
reionization we consider different observational constraints. A first
set includes direct measurements of their high-z LF and its redshift
evolution. In this class, we consider both the optical QSO-LF derived
in the framework of the Great Observatories Origins Deep Survey
(GOODS) collaboration by \citet[F07 hereafter]{Fontanot07a} and the
X-ray selected QSO-LF by \citet{Fiore11}. Moreover, we also consider
the upped limit estimate from \citet{ShankarMathur07}: they use the
results of optical surveys down to the highest redshifts and faintest
magnitudes proven (critically including non-detections) to give
constraints on the faint end of the QSO-LF at $z>5$. Their analysis
shows that the observational constraints found so far are compatible
with a faint-end slope of $\alpha=-2.8$ (at $99\%$ confidence level)
and $\alpha=-2.2$ (at $90\%$ confidence level). We directly compare
these three estimates in Fig.~\ref{fig:qsolf}; X-ray measurements and
the relative analytical fits have been converted into absolute
magnitude $M_{\rm UV}$ by means of the QSO bolometric corrections
proposed by \citet{Marconi04, Elvis94, Fontanot07a}. The three
estimates agree well at the bright-end of the LF; on the other hand
the agreement is considerably reduced at the faint-end of the LF and
at higher redshift due to the different redshift evolution
extrapolated to higher redshift \citep{Fontanot07c}. This behaviour is
in part related to the different selection of QSO databases: it is
indeed quite clear that X-ray surveys observe higher faint QSO space
densities than optical surveys. It is also worth stressing that we
expect obscuration effects to play a relevant role at these
luminosities (see e.g. \citealt{Simpson05}). Moreover, at these
luminosities we are no longer considering only ``bona fide'' QSOs
(i.e. $M_B<-23$), but we are entering a regime where the AGN/QSO
distinction becomes loose.

If the QSO population and its high-z LF is representative only of a
sub-population of the global AGN population, we may as well be missing
a relevant contribution to the ionizing flux.  As an alternative
approach to gain insight into the AGN LF and in particular its faint
end slope we consider the galaxy stellar mass function (GSMF) as
estimated in the redshift range $0.4<z<4.5$ by \citet{Santini12}.
Following simple empirical reasoning we tried to use the constraints
provided by the \citet{Santini12} analysis in their highest redshift
bin ($3.5<z<4.5$). In particular we assume:
\begin{itemize}
\item[(a)]{each galaxy host a Supermassive Black Hole (SMBH) at its
  center, and the relation between galaxy stellar mass and SMBH mass
  is the same as in local Universe \citep[i.e. all galaxies are
    ellipticals]{Magorrian98, HaringRix04}; }
\item[(b)]{the estimated faint end of the GSMF is a good
  representation of the dwarf population;}
\item[(c)]{the high-z QSO-LF is a rescaled version of the GSMF;}
\item[(d)]{at $z>4.5$ the GSMF evolves as the F07 QSO-LF: i.e. an
  exponential pure density evolution proportional to
  $e^{-1.26(1+z)}$};
\item[(e)]{at each redshift only $0.45\%$ of SMBH shine at their
  Eddington luminosity (see e.g. \citealt{Haiman04});}
\item[(f)]{conversion from bolometric luminosity to $M_{\rm UV}$ using
  a standard approach (see e.g., F07): we consider a bolometric
  correction factor for the $B$-band luminosity of 10.4
  \citep{Elvis94} and a $B$- to $UV$-magnitude shift of -0.48 mag.}
\end{itemize}
\noindent
We show the result of these conversions in Fig.~\ref{fig:qsolf}: the
red lines (and hatched light red shading in between) correspond to the
upper and lower best-fits\footnote{In particular, the upper envelope
  corresponds to the best fit to a double power yielding a
  normalization $log(\Phi_\star)=-4.84$, a characteristic mass
  $log(M_\star)=11.81$, an high-mass-end slope $\beta=-6.38$ and a
  low-mass-end slope $\alpha=-2.27$ (table~3 in \citealt{Santini12});
  while the lower envelope refers to the best-fit to a Schechter
  function yielding $log(\Phi_\star)=-4.12$, $log(M_\star)=11.30$ and
  $\alpha=-1.80$ (table~1 in \citealt{Santini12}).} in
\citet{Santini12} in the range $3.5<z<4.5$. It is interesting to
stress that the knee of the ``mass-derived'' luminosity function
results in a good agreement with respect to F07 best-fit models, and
the overall normalization is consistent with the densities obtained in
QSO-LF studies. This is an important sanity check for the proposed
approach, since some of the factors (a) to (f) are degenerate and
uncertain as well.  In particular, there are several evidences in
favour of a redshift evolution of both the ratio between SMBH mass and
the spheroidal host component mass, and on the AGN fraction/duty cycle
(see e.g. \citealt{Shankar09, Fiore11} and discussion herein). The
bright-end of the AGN-LF is steeper than any QSO-LF, while its
faint-end is steeper than both the \citet{Fiore11} and
\citet{Fontanot07a} estimates, but still in the range allowed by the
\citet{ShankarMathur07} analysis.

In order to compute the QSO contribution to reionization, we compute
the rate of emitted ionizing photons per unit comoving volume
$\Gamma_{\rm AGN}$ as a function of redshift, following the same
formalism as in \citet[see also \citealt{MHR99}]{ShankarMathur07}:

\begin{equation}\label{gamma_agn}
\Gamma_{\rm AGN}(z) [\textrm{s}^{-1} \textrm{Mpc}^{-3}] = \int_{\nu_H}^{\nu_{\rm up}} \sigma_\nu
\frac{\rho_\nu(z)}{h_p \nu}
\end{equation}

\begin{equation}\label{lum_den}
\rho_\nu(z) [\textrm{erg s}^{-1} \textrm{Hz}^{-1} \textrm{Mpc}^{-3} ] = \int_{L_{\rm min}}^\infty \Phi(L,z) L_\nu (L) dL
\end{equation}

\noindent
In the above equations, $L_\nu$ is in erg s$^{-1}$ Hz$^{-1}$, $h_p $
represents the Planck's constant, $\nu_H = 3.2 \times 10^{15}$ Hz is
the frequency at the Lyman Limit (i.e. $912$ \AA); $\nu_{\rm up} =
12.8 \times 10^{15}$ Hz is the usual upper limit to the integration,
since photons more energetic are preferentially absorbed by {\rm He}
atoms. In practice, we assume that the absorbing cross section for
neutral hydrogen $\sigma_\nu$ is unity between $\nu_H$ and $\nu_{\rm
  up}$, and zero outside this range. This crude approximation gives us
a good grasp of the maximum contribution of QSO to the reionization
background. We also assume that all the ionizing photons associated
with the AGN spectra contribute to the ionizing background
(i.e. $f_{\rm esc}=1$).  In the calculations, we assume a QSO spectral
continuum of the form $f_{\nu} = \nu^\gamma $ and we assume a slope
$\gamma=-1.76$ blueward of the $Ly_\alpha$ line \citep{Telfer02}. In
the following, we will compare $\Gamma_{\rm AGN}$ with the required
total ionizing photon rate per unit comoving volume $\Gamma_{\rm
  ion}$, using the formalism proposed in \citet{MHR99}, rescaled to
WMAP7 cosmology as in \citep{Pawlik09}:

\begin{equation}
\Gamma_{\rm ion}(z) [\textrm{s}^{-1} \textrm{Mpc}^{-3}] = 0.027 \, \kappa \left(
\frac{C}{30} \right) \left( \frac{1+z}{7} \right)^3 \left(
\frac{\Omega_b h^2_{70}}{0.0465} \right)^2
\end{equation}

\noindent where we convert the critical star formation rate into a
photon rate, by assuming that $\kappa=10^{53.1}$ s$^{-1}$ LyC photons
per M$_\odot$ yr$^{-1}$ are produced (see e.g. \citealt{Shull12}, see
also eq.~\ref{eq:sh} below) and $C$ refers to the clumping factor of
the intergalactic medium. Early work (see e.g.  \citealt{MHR99,
  ShankarMathur07}) assumed a high value $C=30$ for the clumping
factor, following the results of numerical simulations by
\citet{GnedinOstriker97}, and concluding that the space density of
ionizing photons deducted by observations was in most cases
insufficient to reionize the Universe and/or kept it ionized. More
recent theoretical estimates (see e.g.  \citealt{Bolton07,
  Pawlik09,HaardtMadau12} revised the value of the clumping factor
towards lower values. In particular, \citet{Pawlik09} estimates $C=6$
as adequate for gas with densities of the order of the critical
density for the onset of star formation, while finding an even lower
$C=3$ value for gas with overdensities $\sim100$. It is also worth
stressing that $C$ is expected to be a decreasing function of
redshift. For example \citet{HaardtMadau12} propose the following
fitting formula:
  \begin{equation}\label{eq:cz}
    C(z) = 1+43 \times z^{-1.71}
  \end{equation}
\noindent
derived for gas with overdensities $\sim100$. Eq.~\ref{eq:cz} is
fully consistent with the \citet{Pawlik09} estimate on the same
overdensity scale.  Lower values for the clumping factor reduce
considerably the number of ionizing photons required to keep the
Universe ionized at $z>6$, mitigating the requests on the observed
astrophysical sources. In the following, we will assume a redshift
dependent clumping factor as in eq.~\ref{eq:cz}.

\subsection{High-z LBGs Luminosity Functions}

In order to estimate the number of ionizing photons produced by the
galaxy population, we consider the high-z luminosity function of Lyman
Break Galaxies (LBGs hereafter) as estimated by
\citet{Bouwens11a}. Following these authors we describe the LBG-LF in
the redshift range $3.5 \lesssim z \lesssim 8$ as an evolving
Schechter function (i.e. whose parameters are evolving with cosmic
time, see table~1 in \citealt{Bouwens11b}). We then compute the
luminosity density $\rho_{\rm UV)}$ using eq.~\ref{lum_den} and we
convert it to an estimate of the star formation rate density
($\rho_{\rm SFR}$) using the \citet{HaardtMadau12} conversion factor:

\begin{equation}\label{uv2sfr}
\rho_{\rm SFR}(z) [\textrm{M}_\odot \textrm{yr}^{-1} \textrm{Mpc}^{-3}] = \frac{\rho_{\rm UV}(z)
  [\textrm{erg s}^{-1} \textrm{Hz}^{-1} \textrm{Mpc}^{-3}]}{1.05 \times 10^{28}}
\end{equation}

\noindent
Major uncertainties affect the conversion between UV luminosity
density and the $\rho_{\rm SFR}$. First of all, UV observations are
severely affected by dust attenuation. \citet{Bouwens07} estimated
that $z>4$ LBGs suffer lower attenuation levels that lower-z
counterparts, and they suggest that the $\rho_{\rm SFR}$ obtained from
eq.~\ref{uv2sfr} may be underestimated by a factor $\sim
1.5$. Moreover, the conversion factor itself critically depends on the
details of the Stellar Population Synthesis modeling; in particular
the chosen Initial Stellar Mass Function play a relevant role, by
determining the relative abundance of massive stars, the main
contributors to UV fluxes. The value we use in eq.~\ref{uv2sfr} has
been computed assuming a Salpeter IMF. If we consider, i.e. a Kroupa
IMF, its value is reduced by a factor $\sim 1.5$. In the following, we
take into account all these sources of uncertainties by assuming
eq.~\ref{uv2sfr} as a good representation of the mean conversion and
by defining a ``maximal'' and a ``minimal'' model by increasing and
decreasing the resulting $\rho_{\rm SFR}$ by a factor $1.5$,
respectively. We then estimate the rate of ionizing photon production
using the conversion factor proposed for a low-metallicity gas by
\citet[see also \citealt{MHR99}]{Shull12}:

\begin{equation}\label{eq:sh}
\Gamma_{\rm LBG}(z)  [\textrm{s}^{-1} \textrm{Mpc}^{-3}] = \kappa \, f_{\rm esc} \, \rho_{\rm SFR}(z)
\end{equation}

\noindent
where $f_{\rm esc}$ represents the escape fraction\footnote{Most
  observational works focus on the {\it relative} escape fraction,
  i.e. the fraction of escaping LyC photons, relative to the fraction
  of escaping non-ionizing ultraviolet photons. Since the relative
  $f_{\rm esc}$ takes into account the dust attenuation, it is then
  possible to convert between the two determinations. For the purposes
  of this work we just deal with the absolute $f_{\rm esc}$.},
i.e. the fraction of produced ionizing photons which are able to
escape the local environment and ionize the intergalactic medium. This
parameter has a key importance in order to evaluate the contribution
of LBGs to the ionizing background, but its very poor constrained. At
$z \sim 3-4$, proposed values range from low ($f_{\rm esc} < 5\%$,
\citealt{Vanzella10b}), to relatively high values ($f_{\rm esc}
\gtrsim 20\%$, \citealt{Shapley06, Iwata09}).

\section{Discussion}
\label{sec:results}

\begin{figure}
  \centerline{
    \includegraphics[width=9cm]{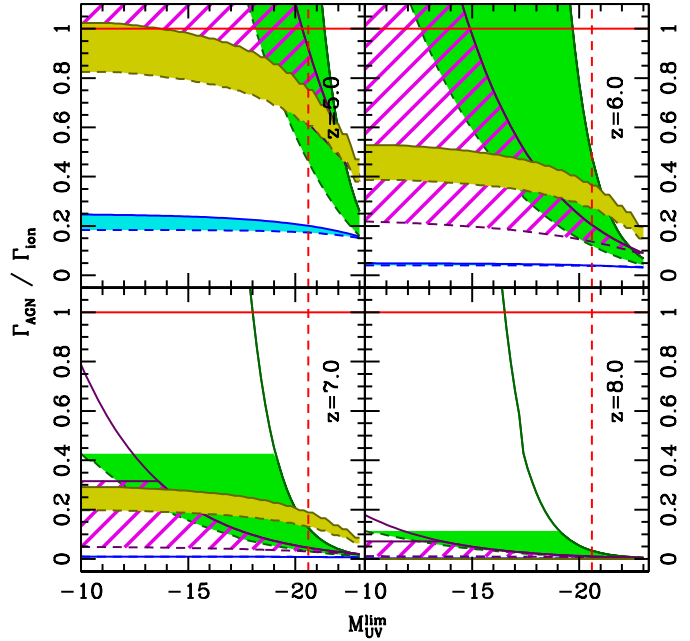}
  }
  \caption{Estimated AGN contribution to reionization at different
    cosmic epochs. Lines and colours as in
    Fig.~\ref{fig:qsolf}. Shaded areas represent allowed contributions
    after Helium reionization has been taken into account (see text
    for more details). Dashed vertical lines represent the actual
    limit ($M_{\rm UV} \sim -20.6$) of faint QSO searches as defined
    in \citet{ShankarMathur07}.}\label{fig:qso}
\end{figure}

\subsection{The AGN contribution to the ionizing background}

First of all we consider the AGN contribution to the ionizing
background. At variance with previous analyses we don't fix $L_{\rm
  min}$ in eq.~\ref{lum_den}, but we study how the ratio $\Gamma_{\rm
  AGN} / \Gamma_{\rm ion}$ evolves as a function of redshift and
limiting magnitude. We also compute the AGN contribution to the
reionization by applying the same approach to our GSMF-derived AGN-LF
(sec.~\ref{subsec:QSO}), and include these results in
Fig.~\ref{fig:qso} (red lines and red hatched area). Following the
same (a) to (f) prescriptions described in sec.~\ref{subsec:QSO}, we
estimate that the putative AGN luminosity associated to a Jeans Mass
of pristine gas\footnote{The limiting mass $M_J$ for the collapse of a
  cloud of cold gas into stars is usually determined by comparing its
  thermal and gravitational energy; by assuming typical values for the
  temperature and density we obtain $M_J \sim 10^5 M_\odot$} is
roughly $M_{\rm UV} \sim -10$. The formation of the ancestors of SMBHs
at very high redshifts has been discussed by a number of authors (see
e.g. \citealt{Petri12} and references herein), showing that both heavy
($>10^5 M_\odot$) and light SMBHs seeds are physically plausible at
the same scales accessible for gas cooling and star formation. We
further assume that all ionizing photons associated with the AGN
spectra are available for ionizing the IGM and we show our results in
Fig.~\ref{fig:qso}.

We find that the QSO contribution at the current observational limit
($M_{\rm UV} \sim -20.6$), is not negligible at $z \sim 6$, but still
insufficient to provide the required rate of ionizing photons.  The
difference with respect to analogous works in the literature
(e.g. \citealt{ShankarMathur07}), is to be ascribed to the lower
clumpiness adopted here.

At higher redshift the contribution of AGNs to the ionizing background
decreases rapidly, becoming of the order of a few percent at $z \sim
8$ and negligible thereafter. To obtain a relevant AGN contribution
(i.e. $>10\%$) at these redshifts a steep faint end is required for
the high-z AGN LF (steeper than the GSMF faint-end slope determined by
\citet{Santini12}) and an integration to very low luminosity limits,
under the hypothesis that very faint AGNs are characterized by the
same properties of their bright counterparts. Another possibility is
that the evolution of the low-luminosity AGN population becomes slower
with respect to the $3<z<5$ estimates.

We have also imposed that our analysis be consistent with present
constraints on the HeII reionization (see
e.g. \citealt{Furlanetto08}), i.e. areas of the parameter space
predicting too many photons with energies larger than 4 Ryd and
therefore producing HeII reionization at redshifts $z \gtrsim 5$ are
forbidden.  To this end we assume the same clumping factor as adopted
for Hydrogen reionization, a ratio between the number densities of
Helium and Hydrogen atoms equal to $Y / 4(1-Y)$ (where $Y=0.258$
represents the cosmic fraction of Helium by mass), a recombination
rate for Helium 6 times faster than for Hydrogen and we integrate our
standard QSO spectral shape between 4 and 16 Ryd. Shaded areas in
fig.~\ref{fig:qso} thus represents the allowed contribution to the
ionizing background according to this analysis.

\begin{figure}
  \centerline{ \includegraphics[width=9cm]{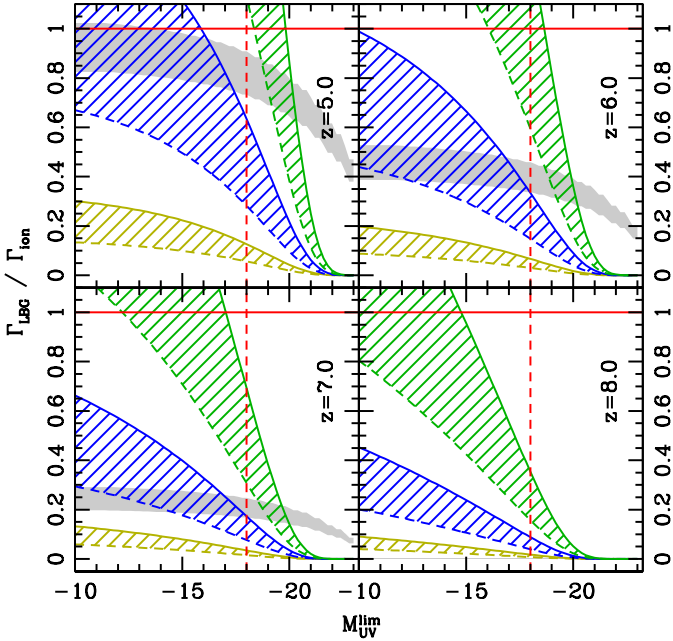} }
  \caption{Estimated LBG contribution to reionization at different
    cosmic epochs. Yellow, blue and green lines and hatched regions
    refer to the integrated contribution of LBGs for $f_{\rm esc}=1\%
    ; 5\% ; 20\%$ respectively. Solid and dashed line refer to the
    ``maximal'' and ``minimal'' models (see text for more
    details). The grey shaded area corresponds to the estimated AGN
    contribution from the \citet{Fiore11} high-z AGN-LF shown in
    Fig.~\ref{fig:qso}. Dashed vertical lines represent the actual
    limit ($M_{\rm UV} \sim -18$) of faint LBG searches as defined in
    \citet{Bouwens11a}. }\label{fig:lbg}
\end{figure}

\subsection{The LBG contribution to the ionizing background}
 
In Fig.~\ref{fig:lbg} we present the estimated contribution of LBGs to
cosmic reionization for different $f_{\rm esc}$ values. Also for this
class of sources we integrate the contribution of sources down to
$M_{\rm UV} \sim -10$. We directly compare these results with the
contribution from the AGN-LF, and in particular with the results of
\citet[][grey shaded area]{Fiore11}. We confirm the result of
\citet{Pawlik09} that observed LBGs may account for the total required
ionizing photon budget at $z\sim6$ if $f_{\rm esc} \sim 20\%$; at
higher redshifts, we have to integrate the LBG-LF to increasingly
fainter limiting magnitudes (up to $M_{\rm UV} \sim -10$ at $z\sim9$),
in order to produce enough ionizing photon to fully account for cosmic
reionization. However, if $f_{\rm esc} \sim 5\%$ for the whole LBG
population, the LBG contribution is not enough to account for the
whole required ionizing photon rate at $z\gtrsim7$, even if we
extrapolate the LBG-LF up to the fainter magnitudes, and it becomes
roughly compatible with that of AGNs for $z\lesssim7$. In order to
achieve high-z reionization with LBGs only, a substantial contribution
to the ionizing background of sources fainter than the actual
observational limit is required and/or a very different
(i.e. increasing) $f_{\rm esc}$ with respect to their bright
counterparts.

In order to test this hypothesis we impose a simple
luminosity-dependent $f_{\rm esc}$ scenario to our maximal model by
defining $f_{\rm esc}=0$ for $M_{\rm UV}<-18.00$ and a linearly
increasing $f_{\rm esc}$ with increasing magnitude:
\begin{equation}\label{eq:ld}
  f_{\rm esc} = \min [1 , \eta \times (M_{\rm UV}+18.00)]
\end{equation}
Despite the lack of constraints on the distribution of $f_{\rm esc}$
in different galaxy population, this exercise allows us to provide a
qualitative estimate for the typical magnitude of objects responsible
for the bulk of reionization and their expected escape fractions, if
$f_{\rm esc}$ is indeed a decreasing function of luminosity. This
simple toy model predicts that complete reionization at $z=8$ by LBGs
alone is achieved at typical magnitudes $M_{\rm UV} \sim -13.5, -15,
-15.5$ for $\eta=0.1, 0.2, 0.3$ respectively. The predicted
corresponding escape fractions are $f_{\rm esc} ~ 50, 60, 80\%$. A
similar result is obtained at $z=9$ (typical magnitudes $M_{\rm UV}
\sim -12.5, -14, -15$ and $f_{\rm esc} \sim 55, 75, 90\%$). This
results are compatible with the recent hydrodynamical simulations (see
e.g., \citealt{Ciardi12}), which require rather high $f_{\rm esc}$
values for fainter LBGs. It is also worth noting that, in general,
lower $f_{\rm esc}$ values are still compatible with an LBG-driven
reionization if a redshift increasing efficiency of ionizing photons
production and/or a top-heavy IMF are assumed (see
e.g. \citealt{Schneider02}).

\begin{figure}
  \centerline{
    \includegraphics[width=9cm]{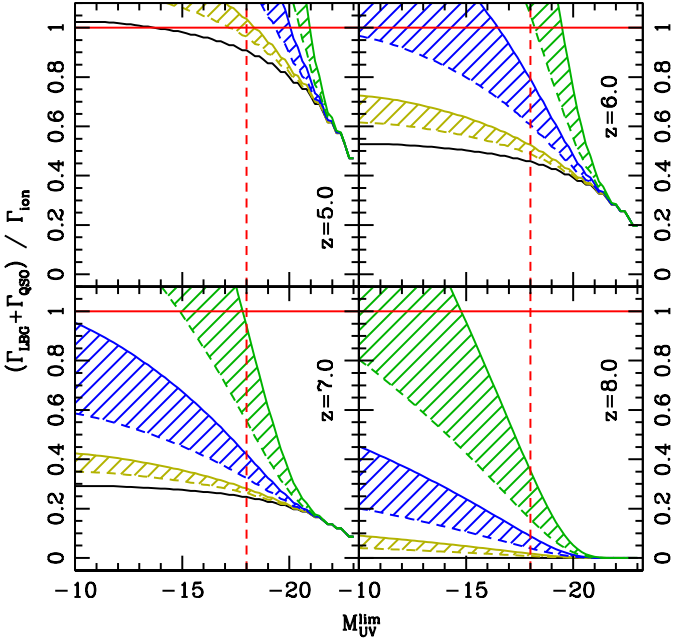}
  }
  \caption{Combined LBG and AGN contributions to reionization. Colours
    and lines as in Fig.~\ref{fig:lbg}. In each panel we consider the
    maximal AGN contribution from the high-z AGN-LF from
    \citet[][marked by the solid line in each panel, see text for more
      details]{Fiore11}.}\label{fig:comb}
\end{figure}

\subsection{Combined AGN-LBG contribution to ionizing background}
Fig.~\ref{fig:comb} shows the combined AGN+LBG contribution to the
reionization of the Universe. For the sake of simplicity we only
consider the maximum AGN contribution as estimated on the basis of the
AGN-LF of \citet[][marked as a solid line in each panel]{Fiore11}:
this is at the same time a conservative but representative choice
among the various LF estimates that we consider in
sec.~\ref{subsec:QSO}.  From a comparison with the results shown in
Fig.~\ref{fig:lbg} it is apparent that AGNs may significantly help
reducing the gap between the LBG ionizing photon production rate and
the required amount of ionizing photons up to $z \sim 7$. For a given
$f_{\rm esc}$ the minimum luminosity of the galaxies required to match
the theoretically estimated ionization limit is significantly
increased. Nonetheless, at the highest redshift considered, it is
still not possible to reach the expected space density of ionizing
photons for reionization if $f_{\rm esc} \sim 5\%$.  Again, in order
to reionize the Universe with LBGs and AGNs combined at such high
redshifts, either $f_{\rm esc}$ has to be higher than the current
estimates (at least for the faint LBG population, which is assumed to
provide the largest contribution), and/or the evolution of the LF has
to slow down with respect to the present estimates.
\begin{figure}
  \centerline{ \includegraphics[width=9cm]{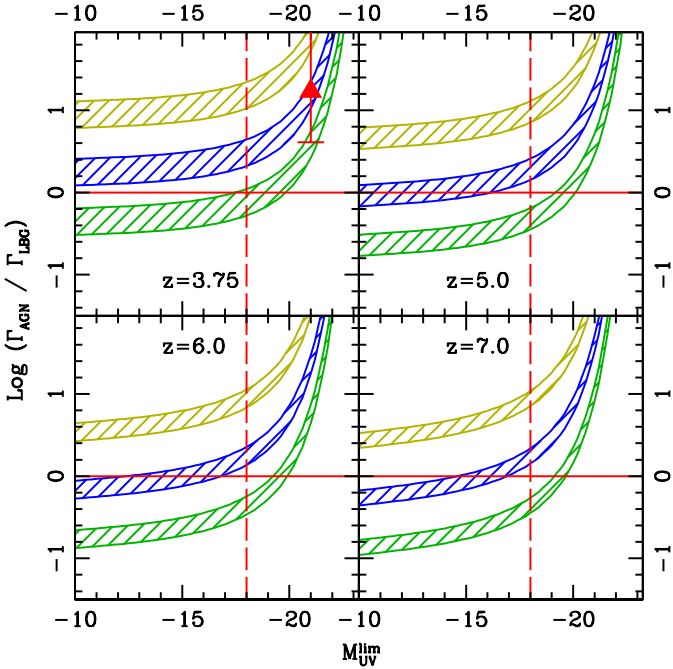} }
  \caption{Logarithmic ratio between the AGN and LBG contributions to
    the reionization. As a reference for AGNs, the \citet{Fiore11}
    AGN-LF has been considered. Colours and lines as in
    Fig.~\ref{fig:lbg}. The red filled triangle marks the position of
    our measurement at $3.4<z<4.0$ carried out in the GOODS fields
    (see text for more details).}\label{fig:ratio}
\end{figure}
In Fig.~\ref{fig:ratio}, we then show the $\Gamma_{\rm
  AGN}/\Gamma_{\rm LBG}$ ratio as a function of $M_{\rm UV}$ and
redshift. For the AGN population we consider as representative the
\citet{Fiore11} LF, while for the LBG population we show results for
$f_{\rm esc}=1\%$, $5\%$ and $20\%$. From Fig.~\ref{fig:ratio} is
clear that $\Gamma_{\rm AGN}$ is decreasing at increasing redshifts at
a faster pace relative to $\Gamma_{\rm LBG}$ independently on $f_{\rm
  esc}$. The latter contribution becomes dominant for faint sources at
$z>5$ for $f_{\rm esc} \gtrsim 5\%$.
\begin{figure}
  \centerline{
\includegraphics[width=9cm]{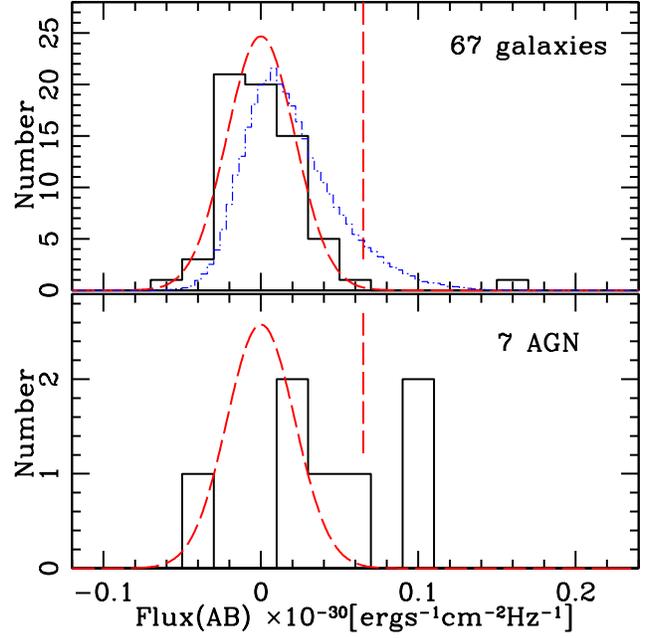} }
  \caption{Distribution of U-band fluxes (probing the rest-frame LyC)
    for 67 LBGs (upper panel) and 7 AGNs (lower panel) in the GOODS
    South field (selected with $3.4 < z < 4$ and $23.5 < i_{775} <
    26$). Fluxes have been measured in both cases within an aperture
    of 1.2'' diameter. The red dashed Gaussian distributions show the
    expected distribution of null detections normalized to the total
    number of measurements in each panel. For comparison, the
    dot-dashed blue histogram shows the expected distribution of
    fluxes for galaxies with an $f_{\rm esc}=10\%$ (the IGM absorption
    being simulated as in \citealt{Vanzella10b}). The vertical dashed
    line marks the $3~\sigma$ confidence limit.}
\label{fig:distribUflux}
\end{figure}

An interesting reference point can be obtained by studying the LBGs
and AGNs in the GOODS South field, selected in equal redshift and
magnitude intervals, $3.4<z<4$ and $23.5<i_{775}<26$. Following the
procedure described in \citet{Vanzella10b}, we have measured the flux
of 67 LBGs and 7 AGN in the U-band (probing the rest-frame LyC) using
a circular aperture of 1.2" diameter. As shown in
Fig.~\ref{fig:distribUflux}, one galaxy (dubbed {\it Ion1} in
\citet{Vanzella12}) and two AGNs are detected above the $3~\sigma$
confidence level. The average flux from AGN turns out to be $(0.041
\pm 0.08) \times 10^{-30}$ erg$^{-1}$ cm$^{-2}$ Hz$^{-1}$. In order to
compute the corresponding quantity for galaxies we have considered
that the distribution of their UV fluxes is characterized by 66
non-detections and one outlier. We have therefore assumed as an
average UV flux of galaxies the flux of the outlier divided by 67 (we
would obtain a similar value by averaging over the whole distribution)
and as $1 ~\sigma$ confidence levels those computed by
\citet{Gehrels86} for small numbers of events (in this case one),
obtaining $(0.0024^{+0.0055}_{-0.0020}) \times 10^{-30}$ erg$^{-1}$
cm$^{-2}$ Hz$^{-1}$. The statistics are poor but allow us to roughly
estimate a ratio between the AGN and LBG contribution to the UV
ionizing background (shown in Fig.~\ref{fig:ratio} with a red
triangle) of $17^{+105}_{-13}$. This value is conveniently independent
on the IGM transmission and is consistent with the $\lesssim 5\%$
estimate of the $f_{\rm esc}$ from galaxies discussed in
\citet{Vanzella10b}.

It is worth noting that a similar measurement and result has been
obtained at lower redshift by \cite{Cowie09}. These authors study the
ionizing fluxes associated with the AGN and galaxy populations at $z
\sim 1.15$ in the GOODS-North field by means of observations with the
Galaxy Evolution Explorer (GALEX). Their results show the presence of
a detectable signal corresponding to known AGNs/QSOs, while stacking
analysis of galaxy images provides no evidence for a significant
ionizing flux (compatible with $f_{\rm esc} \lesssim 1\%$).

\section{Conclusions}
\label{sec:concl}

We have critically discussed in view of recent results the
contribution to cosmic reionization at $z>5$ of both high-z QSOs
\citep{Fontanot07a, ShankarMathur07, Fiore11} and LBGs
\citep{Bouwens11a}. In order to take into account the contribution
from AGN fainter than the current observational depths we have also
used a derivation of the AGN LF based on the evolution of the galaxy
stellar mass function \citep{Santini12}.  In the following we assume
$z \sim 7$ (e.g. \citealt{Mitra12}) as the fiducial redshift for a
rapid transition of the hydrogen from a significantly neutral
condition to a neutral fraction $x_{HI} << 10^{-3}$.

Our results show that
\begin{enumerate}
\item{In order to achieve the HI reionization at $z \sim 7$ the
  properties of the AGN population have to be pushed to rather extreme
  values in terms of steepness of the faint end of the LF ($\alpha
  \lesssim -2$) and contribution of very faint (up to $M_{UV} \sim
  -10$) objects. But in the case such conditions were met we would
  expect the reionization of HeII to take place, owing to the typical
  AGN SED, above redshift $z \sim 5$, which is in contrast with
  present observations (e.g. \citet{Fechner06, Zheng08}). AGN alone,
  at least in their standard manifestation, cannot be responsible for
  the reionization of HI.}
\item{The LBG population may account for the whole photon budget
  needed for reionization, but only if the mean escape fraction of
  this population is of the order of $f_{\rm esc} \sim 20\%$ and very
  faint dwarf galaxies provide a substantial contribution. Despite
  such an high $f_{\rm esc}$ is still compatible with observational
  constraints \citep{Shapley06, Iwata09}, several evidences point out
  that these values are not typical for the whole high-z LBG
  population \citep[see e.g.]{Vanzella10b}. If mean $f_{\rm esc}$
  values are indeed of the order of $5\%$, and more similar to $z\sim
  1$ results \citep{Cowie09}, the contribution of LBGs alone is not
  enough to account for cosmic reionization at $z\gtrsim7$, and we are
  forced either to advocate additional ionizing sources or a strong
  redshift/luminosity evolution of $f_{\rm esc}$.}
\item{If $f_{\rm esc}\sim5\%$, the AGN population can provide a
  significant contribution to the total photon budget and help
  achieving reionization not earlier than $z \sim 7$.}
\item{If $f_{\rm esc}\lesssim5\%$ for the brighter ($L \gtrsim
  L_\star$) LBG galaxies a viable solution is to advocate an
  increasing $f_{\rm esc}$ with decreasing luminosity, as it may be
  expected if shallower potential wells are more easily cleared of
  (neutral) hydrogen gas (see e.g. \citealt{Mori02, Yajima11}). A
  simple toy model with a linear growth of $f_{\rm esc}$ with
  increasing magnitude shows that, in the case of a negligible
  contribution to the ionizing rate from brighter LBGs, we need to
  assume $f_{\rm esc}$ of the order of $70\%$ for sources fainter than
  the present observational limits to provide a substantial
  contribution to reionization. These results are fully consistent
  with the recent findings of \citet{Kuhlen12}: these authors use a
  different approach to compute the contribution of LBGs to the
  ionizing photons, based on the comparison of the LBG-LF with the
  expected Thompson optical depth and observational constraints on the
  Ly$\alpha$ forest, favouring a scenario where $f_{\rm esc}$
  increases from $z=4$ to $z=9$.}
\end{enumerate}
\begin{figure}
  \centerline{ \includegraphics[width=9cm]{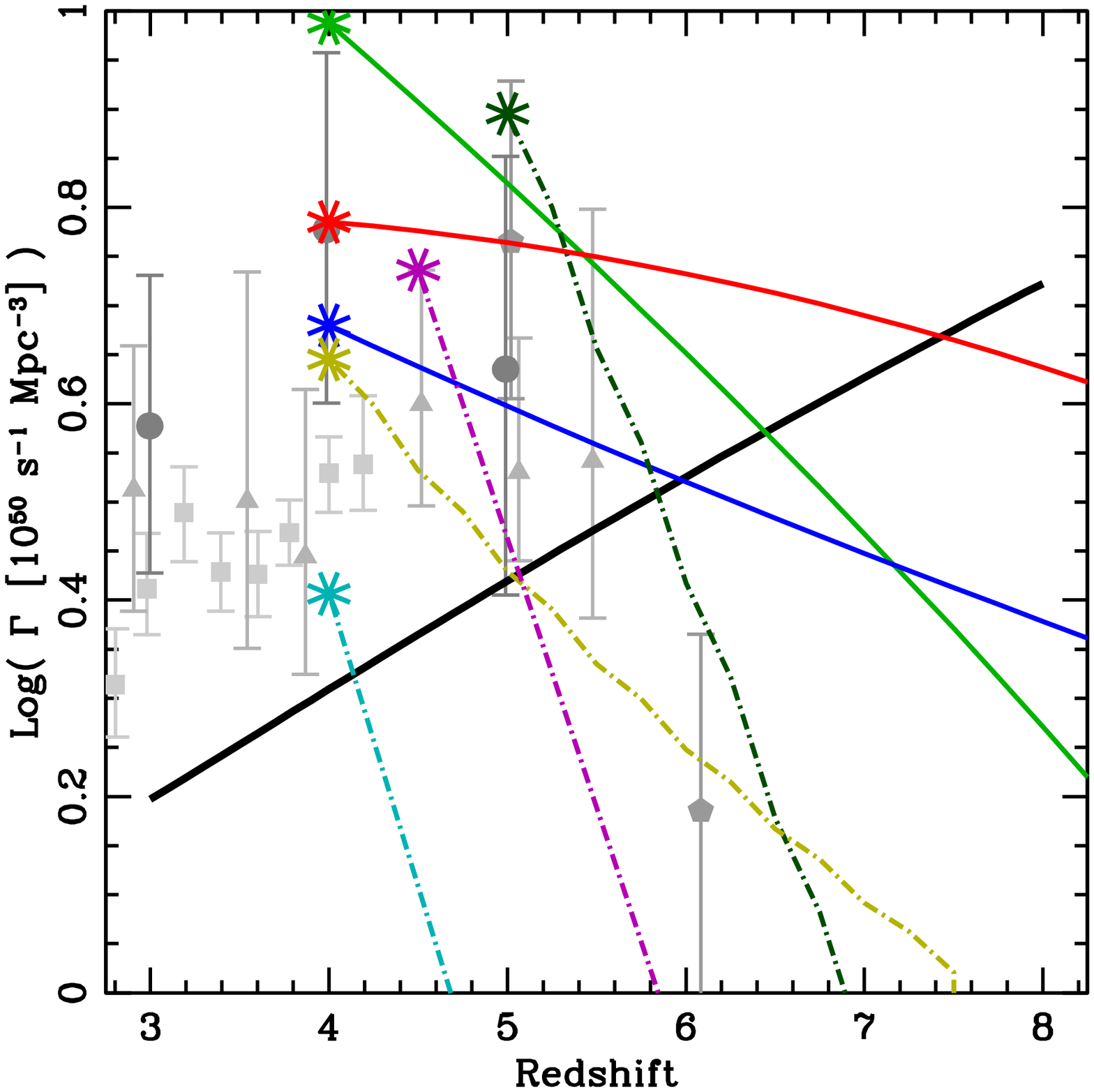} }
  \caption{Comparison between observed and required ionizing
    background. Observed data from \citet[][circles]{Bolton07},
    \citet[][triangles]{Becker07},
    \citet[][squares]{FaucherGiguere08},
    \citet[][pentagons]{Calverley11}. The black solid line mark the
    assumed minimum rate of emitted ionizing photons per unit comoving
    volume for Universe ionization. The cyan asterisk and dot-dashed
    line refers to the level corresponding to the maximum contribution
    from the \citet{Fontanot07a} QSO-LF, integrated up to $M_{\rm
      lim}=-10$; the yellow asterisk and dot-dashed line to the
    maximum contribution from the \citet{Fiore11} QSO-LF, integrated
    up to $M_{\rm lim}=-10$; the dark green asterisk and dot-dashed
    line to maximum contribution from the \citet{ShankarMathur07}
    QSO-LF, integrated up to $M_{\rm lim}=-20$; the dark green
    asterisk and dot-dashed line to maximum contribution from the
    GSMF-derived AGN-LF, integrated up to $M_{\rm lim}=-20$; the blue
    asterisk and solid line to maximum contribution from the
    \citet{Bouwens11b} LBG-LF, integrated up to $M_{\rm lim}=-10$
    assuming $f_{\rm esc}=5\%$; the green asterisk and solid line to
    maximum contribution from the LBG-LF, integrated up to $M_{\rm
      lim}=-18$ assuming $f_{\rm esc}=20\%$; the red asterisk and
    solid line to the luminosity-dependent $f_{\rm esc}$ scenario
    ($\eta=0.1$) integrated up to $M_{\rm lim}=-14$. The different
    slope for the two LBG-LF based models are due to the redshift
    dependent evolution of the LF shape, as modeled by
    \citet{Bouwens11b} results.  }\label{fig:bckgrd}
\end{figure}

In Fig.~\ref{fig:bckgrd} we compare our empirical predictions with the
present observational constraints on the measured ionizing background.
We consider photoionization rate data derived from the observed
Ly$\alpha$-forest effective opacity \citep{Bolton07, Becker07,
  FaucherGiguere08} and from the QSO proximity effect
\citep{Calverley11} and convert them into ionizing photons per unit
comoving volume ($\Gamma_{\rm BKG}$), following the same procedure and
assumptions as in \citet[][their eq. 12]{Kuhlen12}, and using the mean
free path of ionizing photons as measured by \citet{Prochaska09a}. The
assumed minimum rate of emitted ionizing photons per unit comoving
volume for Universe ionization below redshift 5 is only about a factor
2-3 below the measured UV background. In this way, if the sources
responsible for reionization at $z \sim 7$ have a rapid density
evolution, they would quickly exceed the constraint of the measured UV
background.

Fig.~\ref{fig:bckgrd} shows that the space density evolution of AGNs
(as estimated by \citealt{Fontanot07a, ShankarMathur07, Fiore11}) is
too strong for these objects to provide a significant contribution to
reionization at $z\gtrsim7$; LBGs show a milder density evolution for
faint $M_{\rm lim}$ values, steepening as $f_{\rm esc}$ increases (and
$M_{\rm lim}$ brightens), the steepening beeing due to the redshift
dependent evolution of the LF shape \citep{Bouwens11b}.

In both cases our empirical models do not predict enough ionizing
photons at $z>7$, if we force them to obey the constraints on
$\Gamma_{\rm BKG}$ and $f_{\rm esc}$ is kept constant. Also in this
case, the most likely solution for an LBG-driven early reionization
requires an increase of $f_{\rm esc}$ either with decreasing
luminosity (or with increasing redshift): this is clearly shown by the
red line in fig.~\ref{fig:bckgrd}, which represents the contribution
to the ionizing background in the luminosity-dependent $f_{\rm esc}$
scenario (eq.~\ref{eq:ld}, $\eta=0.1$).
  
The exact relative contributions from AGNs and LBGs critically depends
on the details of the slopes of the corresponding faint-end LFs, on
the faintest luminosity limits of these populations and on $f_{\rm
  esc}$. Given the present uncertainties in its determination, better
constraints on the high-z AGN-LF faint-end slope would be of
fundamental importance to understand the maximal contribution of the
AGN population to the ionizing background. In general, our analysis
suggests that pushing the actual observational limits at least one
magnitude fainter, despite very demanding from the observational side,
would be quite rewarding to clearly understand the relative importance
of different astrophysical sources in determining the ionization state
of the early Universe.

\section*{Acknowledgements}
We are grateful to Martin Haehnelt, Andreas Pawlik, Benedetta Ciardi
and Francesco Haardt for enlightening discussions.  FF acknowledges
financial support from the Klaus Tschira Foundation. SC thanks for
stimulating hospitality the Institute of Astronomy of Cambridge.

\bibliographystyle{mn2e}
\bibliography{fontanot}

\label{lastpage}

\end{document}


%% file: qso_reion_rev.bbl
\begin{thebibliography}{}

\bibitem[\protect\citeauthoryear{{Becker}, {Rauch} \& {Sargent}}{{Becker}
  et~al.}{2007}]{Becker07}
{Becker} G.~D.,  {Rauch} M.,    {Sargent} W.~L.~W.,  2007, \apj, 662, 72

\bibitem[\protect\citeauthoryear{{Bolton} \& {Haehnelt}}{{Bolton} \&
  {Haehnelt}}{2007}]{Bolton07}
{Bolton} J.~S.,  {Haehnelt} M.~G.,  2007, \mnras, 382, 325

\bibitem[\protect\citeauthoryear{{Bolton}, {Haehnelt}, {Warren}, {Hewett},
  {Mortlock}, {Venemans}, {McMahon} \& {Simpson}}{{Bolton}
  et~al.}{2011}]{Bolton11}
{Bolton} J.~S.,  {Haehnelt} M.~G.,  {Warren} S.~J.,  {Hewett} P.~C.,
  {Mortlock} D.~J.,  {Venemans} B.~P.,  {McMahon} R.~G.,    {Simpson} C.,
  2011, \mnras, 416, L70

\bibitem[\protect\citeauthoryear{{Bouwens}, {Illingworth}, {Franx} \&
  {Ford}}{{Bouwens} et~al.}{2007}]{Bouwens07}
{Bouwens} R.~J.,  {Illingworth} G.~D.,  {Franx} M.,    {Ford} H.,  2007, \apj,
  670, 928

\bibitem[\protect\citeauthoryear{{Bouwens}, {Illingworth}, {Oesch},
  {Labb{\'e}}, {Trenti}, {van Dokkum}, {Franx}, {Stiavelli}, {Carollo}, {Magee}
  \& {Gonzalez}}{{Bouwens} et~al.}{2011}]{Bouwens11a}
{Bouwens} R.~J.,  {Illingworth} G.~D.,  {Oesch} P.~A.,  {Labb{\'e}} I.,
  {Trenti} M.,  {van Dokkum} P.,  {Franx} M.,  {Stiavelli} M.,  {Carollo}
  C.~M.,  {Magee} D.,    {Gonzalez} V.,  2011, \apj, 737, 90

\bibitem[\protect\citeauthoryear{{Bouwens}, {Illingworth}, {Oesch}, {Trenti},
  {Labbe}, {Franx}, {Stiavelli}, {Carollo}, {van Dokkum} \& {Magee}}{{Bouwens}
  et~al.}{2011}]{Bouwens11b}
{Bouwens} R.~J.,  {Illingworth} G.~D.,  {Oesch} P.~A.,  {Trenti} M.,  {Labbe}
  I.,  {Franx} M.,  {Stiavelli} M.,  {Carollo} C.~M.,  {van Dokkum} P.,
  {Magee} D.,  2011, ArXiv e-prints

\bibitem[\protect\citeauthoryear{{Brusa}, {Civano}, {Comastri}, {Miyaji},
  {Salvato}, {Zamorani}, {Cappelluti} \& {Fiore}}{{Brusa}
  et~al.}{2010}]{Brusa10}
{Brusa} M.,  {Civano} F.,  {Comastri} A.,  {Miyaji} T.,  {Salvato} M.,
  {Zamorani} G.,  {Cappelluti} N.,    {Fiore} F. e.~a.,  2010, \apj, 716, 348

\bibitem[\protect\citeauthoryear{{Calverley}, {Becker}, {Haehnelt} \&
  {Bolton}}{{Calverley} et~al.}{2011}]{Calverley11}
{Calverley} A.~P.,  {Becker} G.~D.,  {Haehnelt} M.~G.,    {Bolton} J.~S.,
  2011, \mnras, 412, 2543

\bibitem[\protect\citeauthoryear{{Ciardi}, {Bolton}, {Maselli} \&
  {Graziani}}{{Ciardi} et~al.}{2011}]{Ciardi12}
{Ciardi} B.,  {Bolton} J.~S.,  {Maselli} A.,    {Graziani} L.,  2011, ArXiv
  e-prints

\bibitem[\protect\citeauthoryear{{Ciardi}, {Ferrara}, {Governato} \&
  {Jenkins}}{{Ciardi} et~al.}{2000}]{Ciardi00}
{Ciardi} B.,  {Ferrara} A.,  {Governato} F.,    {Jenkins} A.,  2000, \mnras,
  314, 611

\bibitem[\protect\citeauthoryear{{Civano}, {Brusa}, {Comastri}, {Elvis},
  {Salvato}, {Zamorani}, {Capak} \& {Fiore}}{{Civano} et~al.}{2011}]{Civano11}
{Civano} F.,  {Brusa} M.,  {Comastri} A.,  {Elvis} M.,  {Salvato} M.,
  {Zamorani} G.,  {Capak} P.,    {Fiore} F. e.~a.,  2011, \apj, 741, 91

\bibitem[\protect\citeauthoryear{{Cowie}, {Barger} \& {Trouille}}{{Cowie}
  et~al.}{2009}]{Cowie09}
{Cowie} L.~L.,  {Barger} A.~J.,    {Trouille} L.,  2009, \apj, 692, 1476

\bibitem[\protect\citeauthoryear{{Elvis}, {Wilkes}, {McDowell}, {Green},
  {Bechtold}, {Willner}, {Oey}, {Polomski} \& {Cutri}}{{Elvis}
  et~al.}{1994}]{Elvis94}
{Elvis} M.,  {Wilkes} B.~J.,  {McDowell} J.~C.,  {Green} R.~F.,  {Bechtold} J.,
   {Willner} S.~P.,  {Oey} M.~S.,  {Polomski} E.,    {Cutri} R.,  1994, \apjs,
  95, 1

\bibitem[\protect\citeauthoryear{{Fan}, {Strauss}, {Becker}, {White}, {Gunn},
  {Knapp}, {Richards}, {Schneider}, {Brinkmann} \& {Fukugita}}{{Fan}
  et~al.}{2006}]{Fan06}
{Fan} X.,  {Strauss} M.~A.,  {Becker} R.~H.,  {White} R.~L.,  {Gunn} J.~E.,
  {Knapp} G.~R.,  {Richards} G.~T.,  {Schneider} D.~P.,  {Brinkmann} J.,
  {Fukugita} M.,  2006, \aj, 132, 117

\bibitem[\protect\citeauthoryear{{Faucher-Gigu{\`e}re}, {Lidz}, {Hernquist} \&
  {Zaldarriaga}}{{Faucher-Gigu{\`e}re} et~al.}{2008}]{FaucherGiguere08}
{Faucher-Gigu{\`e}re} C.-A.,  {Lidz} A.,  {Hernquist} L.,    {Zaldarriaga} M.,
  2008, \apjl, 682, L9

\bibitem[\protect\citeauthoryear{{Fechner}, {Reimers}, {Kriss}, {Baade},
  {Blair}, {Giroux}, {Green}, {Moos}, {Morton}, {Scott}, {Shull}, {Simcoe},
  {Songaila} \& {Zheng}}{{Fechner} et~al.}{2006}]{Fechner06}
{Fechner} C.,  {Reimers} D.,  {Kriss} G.~A.,  {Baade} R.,  {Blair} W.~P.,
  {Giroux} M.~L.,  {Green} R.~F.,  {Moos} H.~W.,  {Morton} D.~C.,  {Scott}
  J.~E.,  {Shull} J.~M.,  {Simcoe} R.,  {Songaila} A.,    {Zheng} W.,  2006,
  \aap, 455, 91

\bibitem[\protect\citeauthoryear{{Fiore}, {Puccetti}, {Grazian}, {Menci},
  {Shankar}, {Santini}, {Piconcelli}, {Koekemoer}, {Fontana}, {Boutsia},
  {Castellano}, {Lamastra}, {Malacaria}, {Feruglio}, {Mathur}, {Miller} \&
  {Pannella}}{{Fiore} et~al.}{2012}]{Fiore11}
{Fiore} F.,  {Puccetti} S.,  {Grazian} A.,  {Menci} N.,  {Shankar} F.,
  {Santini} P.,  {Piconcelli} E.,  {Koekemoer} A.~M.,  {Fontana} A.,  {Boutsia}
  K.,  {Castellano} M.,  {Lamastra} A.,  {Malacaria} C.,  {Feruglio} C.,
  {Mathur} S.,  {Miller} N.,    {Pannella} M.,  2012, \aap, 537, A16

\bibitem[\protect\citeauthoryear{{Fontanot}, {Cristiani}, {Monaco}, {Nonino},
  {Vanzella}, {Brandt}, {Grazian} \& {Mao}}{{Fontanot}
  et~al.}{2007}]{Fontanot07a}
{Fontanot} F.,  {Cristiani} S.,  {Monaco} P.,  {Nonino} M.,  {Vanzella} E.,
  {Brandt} W.~N.,  {Grazian} A.,    {Mao} J.,  2007, \aap, 461, 39

\bibitem[\protect\citeauthoryear{{Fontanot}, {Somerville} \&
  {Jester}}{{Fontanot} et~al.}{2007}]{Fontanot07c}
{Fontanot} F.,  {Somerville} R.~S.,    {Jester} S.,  2007, ArXiv preprint
  (arXiv:0711.1440)

\bibitem[\protect\citeauthoryear{{Furlanetto} \& {Oh}}{{Furlanetto} \&
  {Oh}}{2008}]{Furlanetto08}
{Furlanetto} S.~R.,  {Oh} S.~P.,  2008, \apj, 681, 1

\bibitem[\protect\citeauthoryear{{Gehrels}}{{Gehrels}}{1986}]{Gehrels86}
{Gehrels} N.,  1986, \apj, 303, 336

\bibitem[\protect\citeauthoryear{{Glikman}, {Djorgovski}, {Stern}, {Dey},
  {Jannuzi} \& {Lee}}{{Glikman} et~al.}{2011}]{Glikman11}
{Glikman} E.,  {Djorgovski} S.~G.,  {Stern} D.,  {Dey} A.,  {Jannuzi} B.~T.,
  {Lee} K.-S.,  2011, \apjl, 728, L26+

\bibitem[\protect\citeauthoryear{{Gnedin}, {Kravtsov} \& {Chen}}{{Gnedin}
  et~al.}{2008}]{Gnedin08}
{Gnedin} N.~Y.,  {Kravtsov} A.~V.,    {Chen} H.-W.,  2008, \apj, 672, 765

\bibitem[\protect\citeauthoryear{{Gnedin} \& {Ostriker}}{{Gnedin} \&
  {Ostriker}}{1997}]{GnedinOstriker97}
{Gnedin} N.~Y.,  {Ostriker} J.~P.,  1997, \apj, 486, 581

\bibitem[\protect\citeauthoryear{{Haardt} \& {Madau}}{{Haardt} \&
  {Madau}}{2012}]{HaardtMadau12}
{Haardt} F.,  {Madau} P.,  2012, \apj, 746, 125

\bibitem[\protect\citeauthoryear{{Haiman}, {Ciotti} \& {Ostriker}}{{Haiman}
  et~al.}{2004}]{Haiman04}
{Haiman} Z.,  {Ciotti} L.,    {Ostriker} J.~P.,  2004, \apj, 606, 763

\bibitem[\protect\citeauthoryear{{Haiman} \& {Loeb}}{{Haiman} \&
  {Loeb}}{1998}]{Haiman98}
{Haiman} Z.,  {Loeb} A.,  1998, \apj, 503, 505

\bibitem[\protect\citeauthoryear{{H{\"a}ring} \& {Rix}}{{H{\"a}ring} \&
  {Rix}}{2004}]{HaringRix04}
{H{\"a}ring} N.,  {Rix} H.,  2004, \apjl, 604, L89

\bibitem[\protect\citeauthoryear{{Iwata}, {Inoue}, {Matsuda}, {Furusawa},
  {Hayashino}, {Kousai}, {Akiyama}, {Yamada}, {Burgarella} \&
  {Deharveng}}{{Iwata} et~al.}{2009}]{Iwata09}
{Iwata} I.,  {Inoue} A.~K.,  {Matsuda} Y.,  {Furusawa} H.,  {Hayashino} T.,
  {Kousai} K.,  {Akiyama} M.,  {Yamada} T.,  {Burgarella} D.,    {Deharveng}
  J.-M.,  2009, \apj, 692, 1287

\bibitem[\protect\citeauthoryear{{Jiang}, {Fan}, {Bian}, {Annis}, {Chiu},
  {Jester}, {Lin}, {Lupton}, {Richards}, {Strauss}, {Malanushenko},
  {Malanushenko} \& {Schneider}}{{Jiang} et~al.}{2009}]{Jiang09}
{Jiang} L.,  {Fan} X.,  {Bian} F.,  {Annis} J.,  {Chiu} K.,  {Jester} S.,
  {Lin} H.,  {Lupton} R.~H.,  {Richards} G.~T.,  {Strauss} M.~A.,
  {Malanushenko} V.,  {Malanushenko} E.,    {Schneider} D.~P.,  2009, \aj, 138,
  305

\bibitem[\protect\citeauthoryear{{Komatsu}, {Smith}, {Dunkley}, {Bennett},
  {Gold} et~al.,}{{Komatsu} et~al.}{2011}]{WMAP7}
{Komatsu} E.,  {Smith} K.~M.,  {Dunkley} J.,  {Bennett} C.~L.,  {Gold} B.,
  et~al., 2011, \apjs, 192, 18

\bibitem[\protect\citeauthoryear{{Krumholz} \& {Dekel}}{{Krumholz} \&
  {Dekel}}{2011}]{KrumholzDekel11}
{Krumholz} M.~R.,  {Dekel} A.,  2011, ArXiv e-prints

\bibitem[\protect\citeauthoryear{{Kuhlen} \& {Faucher-Giguere}}{{Kuhlen} \&
  {Faucher-Giguere}}{2012}]{Kuhlen12}
{Kuhlen} M.,  {Faucher-Giguere} C.-A.,  2012, ArXiv e-prints

\bibitem[\protect\citeauthoryear{{Kuhlen}, {Krumholz}, {Madau}, {Smith} \&
  {Wise}}{{Kuhlen} et~al.}{2012}]{Kuhlen11}
{Kuhlen} M.,  {Krumholz} M.~R.,  {Madau} P.,  {Smith} B.~D.,    {Wise} J.,
  2012, \apj, 749, 36

\bibitem[\protect\citeauthoryear{{Madau}, {Haardt} \& {Rees}}{{Madau}
  et~al.}{1999}]{MHR99}
{Madau} P.,  {Haardt} F.,    {Rees} M.~J.,  1999, \apj, 514, 648

\bibitem[\protect\citeauthoryear{{Madau}, {Rees}, {Volonteri}, {Haardt} \&
  {Oh}}{{Madau} et~al.}{2004}]{Madau04}
{Madau} P.,  {Rees} M.~J.,  {Volonteri} M.,  {Haardt} F.,    {Oh} S.~P.,  2004,
  \apj, 604, 484

\bibitem[\protect\citeauthoryear{{Magorrian}, {Tremaine}, {Richstone},
  {Bender}, {Bower}, {Dressler}, {Faber}, {Gebhardt}, {Green}, {Grillmair},
  {Kormendy} \& {Lauer}}{{Magorrian} et~al.}{1998}]{Magorrian98}
{Magorrian} J.,  {Tremaine} S.,  {Richstone} D.,  {Bender} R.,  {Bower} G.,
  {Dressler} A.,  {Faber} S.~M.,  {Gebhardt} K.,  {Green} R.,  {Grillmair} C.,
  {Kormendy} J.,    {Lauer} T.,  1998, \aj, 115, 2285

\bibitem[\protect\citeauthoryear{{Marconi}, {Risaliti}, {Gilli}, {Hunt},
  {Maiolino} \& {Salvati}}{{Marconi} et~al.}{2004}]{Marconi04}
{Marconi} A.,  {Risaliti} G.,  {Gilli} R.,  {Hunt} L.~K.,  {Maiolino} R.,
  {Salvati} M.,  2004, \mnras, 351, 169

\bibitem[\protect\citeauthoryear{{McGreer}, {Mesinger} \& {Fan}}{{McGreer}
  et~al.}{2011}]{McGreer11}
{McGreer} I.~D.,  {Mesinger} A.,    {Fan} X.,  2011, \mnras, 415, 3237

\bibitem[\protect\citeauthoryear{{Mitra}, {Choudhury} \& {Ferrara}}{{Mitra}
  et~al.}{2012}]{Mitra12}
{Mitra} S.,  {Choudhury} T.~R.,    {Ferrara} A.,  2012, \mnras, 419, 1480

\bibitem[\protect\citeauthoryear{{Mori}, {Ferrara} \& {Madau}}{{Mori}
  et~al.}{2002}]{Mori02}
{Mori} M.,  {Ferrara} A.,    {Madau} P.,  2002, \apj, 571, 40

\bibitem[\protect\citeauthoryear{{Ono}, {Ouchi}, {Mobasher}, {Dickinson},
  {Penner}, {Shimasaku}, {Weiner}, {Kartaltepe}, {Nakajima}, {Nayyeri},
  {Stern}, {Kashikawa} \& {Spinrad}}{{Ono} et~al.}{2012}]{Ono12}
{Ono} Y.,  {Ouchi} M.,  {Mobasher} B.,  {Dickinson} M.,  {Penner} K.,
  {Shimasaku} K.,  {Weiner} B.~J.,  {Kartaltepe} J.~S.,  {Nakajima} K.,
  {Nayyeri} H.,  {Stern} D.,  {Kashikawa} N.,    {Spinrad} H.,  2012, \apj,
  744, 83

\bibitem[\protect\citeauthoryear{{Pawlik}, {Schaye} \& {van
  Scherpenzeel}}{{Pawlik} et~al.}{2009}]{Pawlik09}
{Pawlik} A.~H.,  {Schaye} J.,    {van Scherpenzeel} E.,  2009, \mnras, 394,
  1812

\bibitem[\protect\citeauthoryear{{Pentericci}, {Fontana}, {Vanzella},
  {Castellano}, {Grazian}, {Dijkstra}, {Boutsia}, {Cristiani}, {Dickinson},
  {Giallongo}, {Giavalisco}, {Maiolino}, {Moorwood}, {Paris} \&
  {Santini}}{{Pentericci} et~al.}{2011}]{Pentericci11}
{Pentericci} L.,  {Fontana} A.,  {Vanzella} E.,  {Castellano} M.,  {Grazian}
  A.,  {Dijkstra} M.,  {Boutsia} K.,  {Cristiani} S.,  {Dickinson} M.,
  {Giallongo} E.,  {Giavalisco} M.,  {Maiolino} R.,  {Moorwood} A.,  {Paris}
  D.,    {Santini} P.,  2011, \apj, 743, 132

\bibitem[\protect\citeauthoryear{{Petri}, {Ferrara} \& {Salvaterra}}{{Petri}
  et~al.}{2012}]{Petri12}
{Petri} A.,  {Ferrara} A.,    {Salvaterra} R.,  2012, \mnras, 422, 1690

\bibitem[\protect\citeauthoryear{{Pierpaoli}}{{Pierpaoli}}{2004}]{Pierpaoli04}
{Pierpaoli} E.,  2004, Physical Review Letters, 92, 031301

\bibitem[\protect\citeauthoryear{{Prochaska}, {Worseck} \&
  {O'Meara}}{{Prochaska} et~al.}{2009}]{Prochaska09a}
{Prochaska} J.~X.,  {Worseck} G.,    {O'Meara} J.~M.,  2009, \apjl, 705, L113

\bibitem[\protect\citeauthoryear{{Robertson}, {Ellis}, {Dunlop}, {McLure} \&
  {Stark}}{{Robertson} et~al.}{2010}]{Robertson10}
{Robertson} B.~E.,  {Ellis} R.~S.,  {Dunlop} J.~S.,  {McLure} R.~J.,    {Stark}
  D.~P.,  2010, \nat, 468, 49

\bibitem[\protect\citeauthoryear{{Santini}, {Fontana}, {Grazian}, {Salimbeni},
  {Fontanot}, {Paris}, {Boutsia}, {Castellano}, {Fiore}, {Gallozzi},
  {Giallongo}, {Koekemoer}, {Menci}, {Pentericci} \& {Somerville}}{{Santini}
  et~al.}{2012}]{Santini12}
{Santini} P.,  {Fontana} A.,  {Grazian} A.,  {Salimbeni} S.,  {Fontanot} F.,
  {Paris} D.,  {Boutsia} K.,  {Castellano} M.,  {Fiore} F.,  {Gallozzi} S.,
  {Giallongo} E.,  {Koekemoer} A.~M.,  {Menci} N.,  {Pentericci} L.,
  {Somerville} R.~S.,  2012, \aap, 538, A33

\bibitem[\protect\citeauthoryear{{Schenker}, {Stark}, {Ellis}, {Robertson},
  {Dunlop}, {McLure}, {Kneib} \& {Richard}}{{Schenker}
  et~al.}{2012}]{Schenker12}
{Schenker} M.~A.,  {Stark} D.~P.,  {Ellis} R.~S.,  {Robertson} B.~E.,  {Dunlop}
  J.~S.,  {McLure} R.~J.,  {Kneib} J.-P.,    {Richard} J.,  2012, \apj, 744,
  179

\bibitem[\protect\citeauthoryear{{Schirber} \& {Bullock}}{{Schirber} \&
  {Bullock}}{2003}]{SchirberBullock03}
{Schirber} M.,  {Bullock} J.~S.,  2003, \apj, 584, 110

\bibitem[\protect\citeauthoryear{{Schneider}, {Ferrara}, {Natarajan} \&
  {Omukai}}{{Schneider} et~al.}{2002}]{Schneider02}
{Schneider} R.,  {Ferrara} A.,  {Natarajan} P.,    {Omukai} K.,  2002, \apj,
  571, 30

\bibitem[\protect\citeauthoryear{{Scott}, {Rees} \& {Sciama}}{{Scott}
  et~al.}{1991}]{Scott91}
{Scott} D.,  {Rees} M.~J.,    {Sciama} D.~W.,  1991, \aap, 250, 295

\bibitem[\protect\citeauthoryear{{Shankar}}{{Shankar}}{2009}]{Shankar09}
{Shankar} F.,  2009, \nar, 53, 57

\bibitem[\protect\citeauthoryear{{Shankar} \& {Mathur}}{{Shankar} \&
  {Mathur}}{2007}]{ShankarMathur07}
{Shankar} F.,  {Mathur} S.,  2007, \apj, 660, 1051

\bibitem[\protect\citeauthoryear{{Shapley}, {Steidel}, {Pettini}, {Adelberger}
  \& {Erb}}{{Shapley} et~al.}{2006}]{Shapley06}
{Shapley} A.~E.,  {Steidel} C.~C.,  {Pettini} M.,  {Adelberger} K.~L.,    {Erb}
  D.~K.,  2006, \apj, 651, 688

\bibitem[\protect\citeauthoryear{{Shull}, {Harness}, {Trenti} \&
  {Smith}}{{Shull} et~al.}{2012}]{Shull12}
{Shull} J.~M.,  {Harness} A.,  {Trenti} M.,    {Smith} B.~D.,  2012, \apj, 747,
  100

\bibitem[\protect\citeauthoryear{{Siana}, {Teplitz}, {Ferguson}, {Brown},
  {Giavalisco}, {Dickinson}, {Chary}, {de Mello}, {Conselice}, {Bridge},
  {Gardner}, {Colbert} \& {Scarlata}}{{Siana} et~al.}{2010}]{Siana10}
{Siana} B.,  {Teplitz} H.~I.,  {Ferguson} H.~C.,  {Brown} T.~M.,  {Giavalisco}
  M.,  {Dickinson} M.,  {Chary} R.-R.,  {de Mello} D.~F.,  {Conselice} C.~J.,
  {Bridge} C.~R.,  {Gardner} J.~P.,  {Colbert} J.~W.,    {Scarlata} C.,  2010,
  \apj, 723, 241

\bibitem[\protect\citeauthoryear{{Simpson}}{{Simpson}}{2005}]{Simpson05}
{Simpson} C.,  2005, \mnras, 360, 565

\bibitem[\protect\citeauthoryear{{Telfer}, {Zheng}, {Kriss} \&
  {Davidsen}}{{Telfer} et~al.}{2002}]{Telfer02}
{Telfer} R.~C.,  {Zheng} W.,  {Kriss} G.~A.,    {Davidsen} A.~F.,  2002, \apj,
  565, 773

\bibitem[\protect\citeauthoryear{{Trenti}, {Stiavelli}, {Bouwens}, {Oesch},
  {Shull}, {Illingworth}, {Bradley} \& {Carollo}}{{Trenti}
  et~al.}{2010}]{Trenti10}
{Trenti} M.,  {Stiavelli} M.,  {Bouwens} R.~J.,  {Oesch} P.,  {Shull} J.~M.,
  {Illingworth} G.~D.,  {Bradley} L.~D.,    {Carollo} C.~M.,  2010, \apjl, 714,
  L202

\bibitem[\protect\citeauthoryear{{Vanzella}, {Giavalisco}, {Inoue}, {Nonino},
  {Fontanot}, {Cristiani}, {Grazian}, {Dickinson}, {Stern}, {Tozzi},
  {Giallongo}, {Ferguson}, {Spinrad}, {Boutsia}, {Fontana}, {Rosati} \&
  {Pentericci}}{{Vanzella} et~al.}{2010b}]{Vanzella10b}
{Vanzella} E.,  {Giavalisco} M.,  {Inoue} A.~K.,  {Nonino} M.,  {Fontanot} F.,
  {Cristiani} S.,  {Grazian} A.,  {Dickinson} M.,  {Stern} D.,  {Tozzi} P.,
  {Giallongo} E.,  {Ferguson} H.,  {Spinrad} H.,  {Boutsia} K.,  {Fontana} A.,
  {Rosati} P.,    {Pentericci} L.,  2010b, \apj, 725, 1011

\bibitem[\protect\citeauthoryear{{Vanzella}, {Guo}, {Giavalisco}, {Grazian},
  {Castellano}, {Cristiani}, {Dickinson} \& {Fontana}}{{Vanzella}
  et~al.}{2012}]{Vanzella12}
{Vanzella} E.,  {Guo} Y.,  {Giavalisco} M.,  {Grazian} A.,  {Castellano} M.,
  {Cristiani} S.,  {Dickinson} M.,    {Fontana} A. e.~a.,  2012, \apj, 751, 70

\bibitem[\protect\citeauthoryear{{Vanzella}, {Siana}, {Cristiani} \&
  {Nonino}}{{Vanzella} et~al.}{2010a}]{Vanzella10a}
{Vanzella} E.,  {Siana} B.,  {Cristiani} S.,    {Nonino} M.,  2010a, \mnras,
  404, 1672

\bibitem[\protect\citeauthoryear{{Yajima}, {Choi} \& {Nagamine}}{{Yajima}
  et~al.}{2011}]{Yajima11}
{Yajima} H.,  {Choi} J.-H.,    {Nagamine} K.,  2011, \mnras, 412, 411

\bibitem[\protect\citeauthoryear{{Zahn}, {Reichardt}, {Shaw}, {Lidz}, {Aird},
  {Benson}, {Bleem} \& {Carlstrom}}{{Zahn} et~al.}{2011}]{Zahn11}
{Zahn} O.,  {Reichardt} C.~L.,  {Shaw} L.,  {Lidz} A.,  {Aird} K.~A.,  {Benson}
  B.~A.,  {Bleem} L.~E.,    {Carlstrom} J.~E. e.~a.,  2011, ArXiv e-prints

\bibitem[\protect\citeauthoryear{{Zheng}, {Meiksin}, {Pifko}, {Anderson},
  {Hogan}, {Tittley}, {Kriss}, {Chiu}, {Schneider}, {York} \&
  {Weinberg}}{{Zheng} et~al.}{2008}]{Zheng08}
{Zheng} W.,  {Meiksin} A.,  {Pifko} K.,  {Anderson} S.~F.,  {Hogan} C.~J.,
  {Tittley} E.,  {Kriss} G.~A.,  {Chiu} K.,  {Schneider} D.~P.,  {York} D.~G.,
    {Weinberg} D.~H.,  2008, \apj, 686, 195

\end{thebibliography}
